\documentclass[preprint,showpacs,preprintnumbers,amsmath,amssymb]{revtex4}
\usepackage{psfrag}
\usepackage{graphicx}
\usepackage{dcolumn}
\usepackage{bm}

\begin{document}

\title{To knot or not to knot}

\author{J. Hickford$^1$, R. Jones$^1$, S. Courrech du Pont$^{2,3}$, 
and J. Eggers$^2$ }

\affiliation{
$^1$ H.H. Wills Physics Laboratory, University of Bristol \\
Bristol BS8 1TL, United Kingdom
$^2$School of Mathematics, 
University of Bristol, University Walk, \\
Bristol BS8 1TW, United Kingdom 
$^3$present address: Ecole Normale Sup\'erieure, \\
Laboratoire Mati\`ere et Syst\`emes Complexes, \\
24, rue Lhomond, 75005 Paris (France)
}

\begin{abstract}
We study the formation of knots on a macroscopic ball-chain,
which is shaken on a horizontal plate at 12 times the acceleration 
of gravity. We find that above a certain critical length,
the knotting probability is independent of chain length, while 
the time to shake out a knot increases rapidly with chain length.
The probability if finding a knot after a certain time is the result
of the balance of these two processes. In particular, the knotting
probability tends to a constant for long chains. 
\end{abstract}

\maketitle
\section{Introduction} 
Knots are prevalent on most cables, chains, and strings being
used in every day life or technology. Most remarkably,
knots appear to form 
{\it spontaneously}, as soon as strings are shaken, transported, 
or handled in any way, and thus are an unavoidable byproduct
of their use. For example,
\cite{B01} describe the dynamics of a ball chain that is suspended
from an oscillating support. As soon as the chain dynamics become
chaotic, the chain forms knots of various kinds. Yet in spite
of a considerable amount of work on the importance of knots on
the molecular scale \cite{WC86,KM91}, we are not aware of any systematic 
study into the origin of the prevalence of knots on macroscopic 
chains.

In this paper we present model experiments on shaken ball chains 
that quantify the tendency for knot formation as function of chain
length. By considering both knotting and unknotting events, we present
a simple theory that explains the probability for the formation of 
knots after the chain has been shaken for a given amount of time. 
After the experiment, the ends of the chain lie flat on the plate,
and there is a unique way to join the ends to form a closed curve. 
If this curve is topologically equivalent to a closed loop or 
``unknot'' \cite{A00} there is no knot, otherwise we call the
chain knotted. We made no distinction between different kinds of 
knots, however the simple trefoil knot \cite{A00} was by far the 
most common. 

Of course, it is precisely the topological stability of knots that 
lies at the root of the phenomenon: once a knot is created, it cannot
disappear, except when it falls out at the end of the chain. 
Using a setup very similar to ours, the pioneering study \cite{BDVE01}
investigated the lifetime of a simple trefoil knot that was placed 
in the middle of the chain at the beginning of the experiment. 
The mean lifetime $\tau$ of the knot was found to increase rapidly 
with chain length: 
\begin{equation}
\tau=\tau_3(N-N_0)^2/D,
\label{BN}
\end{equation}
where N is the number of balls on a chain, $N_0\approx 15$ the size of a 
knot, and $D$ a hopping rate. The knot was modeled by the three 
points of intersection of the chain, which perform random walks.
From this assumption the constant $\tau_3=0.056213$, as well as 
the entire distribution of lifetimes was calculated. The experimentally
determined hopping rate of $D=11\pm 1 s^{-1}$ was found to be close 
to the driving frequency of $f=13 Hz$; however, it has not been 
investigated systematically what sets this time scale. 

\begin{figure}
\includegraphics[width=0.7\hsize]{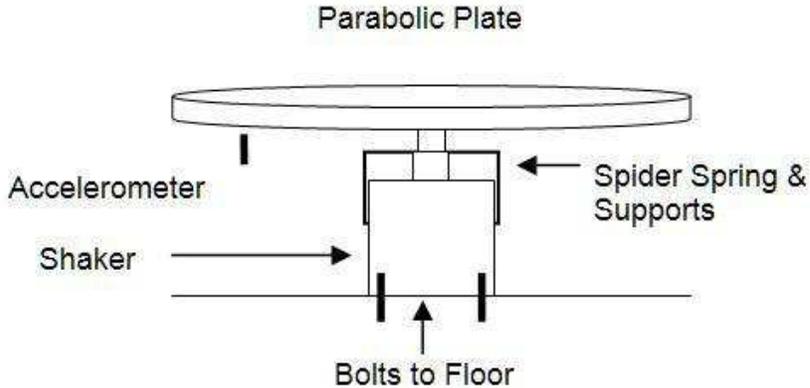}
\caption{Sketch of the experimental setup.\label{apparatus}}
\end{figure}

In Fig. \ref{apparatus} we present a sketch of our experimental setup.
A stiff solid plate, 50 cm in diameter, is vibrated vertically by an 
electrodynamic shaker. The plate's weight of 4 kg is sufficiently 
large for the chains to have little effect on the motion of the plate.
The plate is attached to a metal block mounted on a spider 
spring, to support the weight of the plate and to reduce 
lateral motion. The shaker is controlled by a computer, with an
accelerometer providing feedback to make sure the plate motion 
is close to sinusoidal. The driving frequency $f= 21 Hz$ is dictated 
by the requirement of operating close to resonance, the dimensionless
acceleration is $\Gamma=A\omega^2/g=12\pm 0.2$ as recorded by the 
accelerometer. Here $A$ is the amplitude, $\omega=2\pi f$ the angular
frequency, and $g$ the acceleration of gravity. The top of the plate 
was machined to have a very shallow parabolic profile, of 5mm depth at
the center of the plate. This amount of confinement was enough to 
always keep the chains near the center, without them ever feeling
the edge of the plate. A digital camera was mounted above the plate,
capable of taking up to 20 frames per second. 

Ball chains are an excellent model system to study knot formation
\cite{BDVE01}, in that they have little stiffness that would
resist the formation of loops, yet considerable friction between the 
beads keeps knots from opening too easily. All our chains were cut
from a single sample that had a bead diameter of 
$2R_{bead}=2.1\pm 0.05 mm$, with connecting rods that allowed for 
a maximum inter-bead spacing of $2.1\pm 0.05 mm=2R_{bead}$, somewhat
greater than that of the earlier study ($0.8 R_{bead}$) \cite{BDVE01}.
It seems reasonable to take the number of beads as the fundamental unit
of length, since they offer the greatest resistance as two parts of the 
chain slide across each other. 

\section{Experimental results} 
\begin{figure}
\includegraphics[width=0.7\hsize]{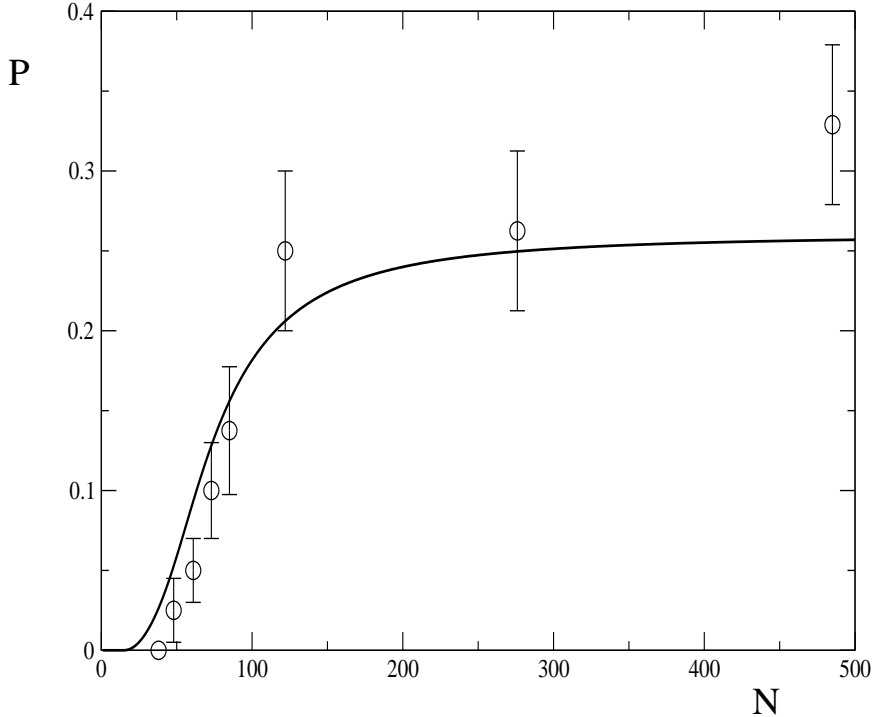}
\caption{Probability for knotting after 30 seconds. 
Error bars are calculated from the variance of a binomial distribution. 
The full line is based on equation (\ref{p}) below. 
\label{comparison}}
\end{figure}

Our main aim is to understand and to quantify the tendency for knots 
to form spontaneously once excited by shaking. To that end we 
shake chains of lengths between N=10 and 500 by dropping them 
onto the vibrating plate, and inspecting them for knots after 30 seconds.
Each experiment was repeated 80 times, and the resulting probability of
knotting calculated, as shown in Fig. \ref{comparison}. No knotting 
was ever observed for chains shorter than $N_{min} = 38$. The probability
then rises sharply to reach a plateau value of about $P=0.26$. 

\begin{figure}
\includegraphics[width=0.7\hsize]{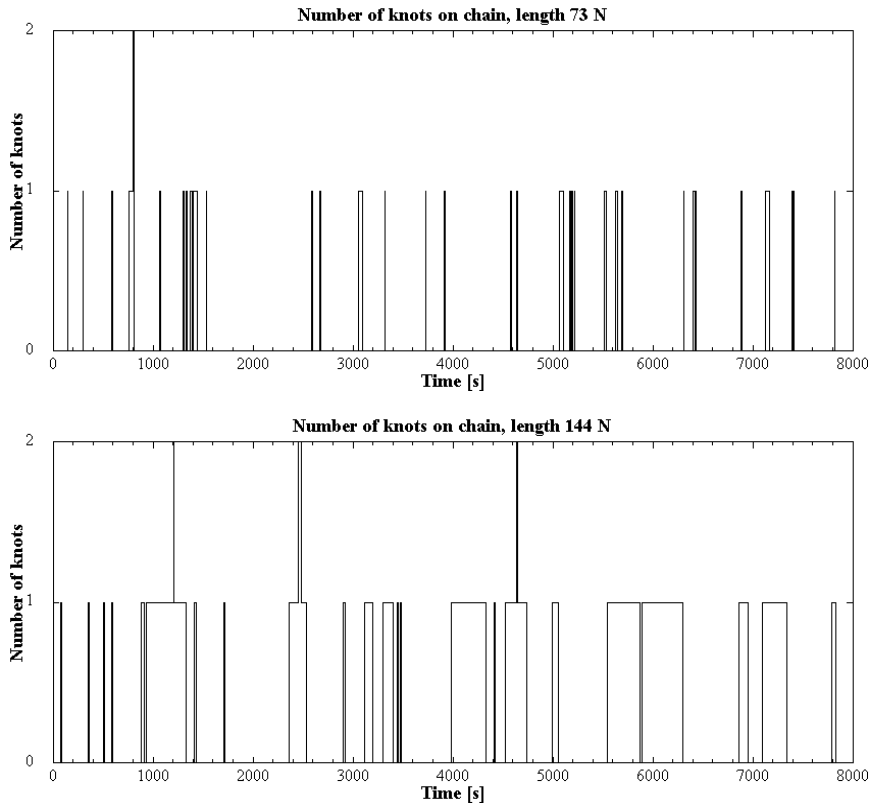}
\caption{The number of knots on chains of length N=73 and N=144 
as a function of time.
\label{evolution}}
\end{figure}

We expect that Fig. \ref{comparison} can be understood from the 
interplay of knotting and unknotting events. To investigate this 
further, we followed the evolution of chains whose length
lay in the transition region (N=73 and N=144) continuously for two hours. 
By taking 10 frames at 20 fps we were able to decide unambiguously 
whether a knot was present. This process was repeated every 
5.5 seconds, which permitted us to compress and store the video images.
The entire sequence was then examined manually for the number of knots,
separated by a length of chain. As shown in Fig. \ref{evolution}, the presence
of two knots is still quite unlikely for the chain lengths shown here. In 
calculating the statistics of knotting and unknotting, knots were 
treated as independent. 

Some statistics as extracted from Fig.\ref{evolution} are collected in 
Table \ref{t1}. As expected, the lifetime of a knot increases 
considerably with length. Comparing to the results of \cite{BDVE01},
there is remarkably good agreement, using (\ref{BN}) with the same 
value of the hopping rate $D$ as in the original paper. This is 
perhaps surprising, considering the differences in chain properties, driving 
conditions and, in particular, in the way knots are introduced. 
In \cite{BDVE01}, knots were introduced by hand in the center of the
chain, while spontaneous knots tend to originate from the end of the 
chain, see Fig. \ref{knot} below. Note that our driving frequency was
also somewhat larger than that of \cite{BDVE01}.
\begin{table}
  \begin{center} 
    \leavevmode 
\begin{tabular}{ccccccccc} 
Chain length [N] & 73 & 144 & \\ \hline 
Chain length [cm] & 30 & 60 & \\ 
mean knot lifetime [s] & $13\pm 5$ & $102\pm 29$ & \\ 
$\tau$ [s] \cite{BDVE01} & 17 & 85 & \\ 
mean knotting time [s] & $219\pm 12$ & $219\pm 16$ & \\ 
$<R_g>$ [cm] (no knot)& $7.18\pm0.03$ & $11.58\pm 0.06$ & \\ 
$<R_g>$ [cm] (knot)& $6.39\pm0.12$ & $11.00\pm 0.09$ & \\ 
\end{tabular}  
  \end{center} 
\caption{Some statistics of knotting and unknotting
for two chains, N=73 and N=144. The last two lines report
the mean radius of gyration.
    } 
\label{t1}
\end{table} 

\begin{figure}
\psfrag{N}{\large $K$}
\psfrag{t}{\large $t [s]$}
\psfrag{n1}{\large $K=K_{tot}(1-\exp(-t/\tau_k))$}
\psfrag{n2}{\large $\tau_k=219 s$}
\includegraphics[width=0.7\hsize]{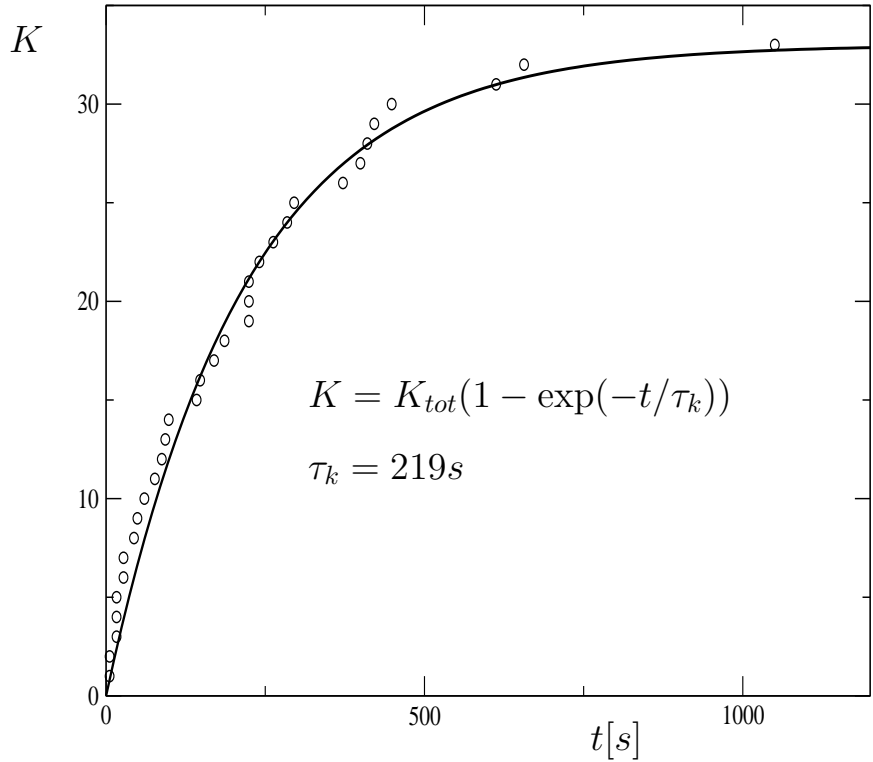}
\caption{
Cumulative number $K$ of knotting events occurring after a shaking time 
of $t$ or smaller. 
\label{total}}
\end{figure}

However the most remarkable observation is that the mean 
knotting time for the two chains is {\it the same},
although the chain lengths are quite different. This observation
makes sense, since knots are produced by the 
ends of the chain, which have sufficient freedom of motion to
wrap around the rest of the chain, as illustrated by the three
knotting events shown in Fig. \ref{knot}. Of course, there is 
certain minimum length that is required, which according to 
Fig. \ref{comparison} is $N_{min}=38$, about twice the minimum
size $N_0$ of a knot. Our data for $N=73$ indicates that the 
knotting rate rises very quickly to a plateau value in the chain 
length. We will assume that the knotting rate is actually constant
once $N$ is larger than $N_{min}$. 
\begin{figure}
\includegraphics[width=0.7\hsize]{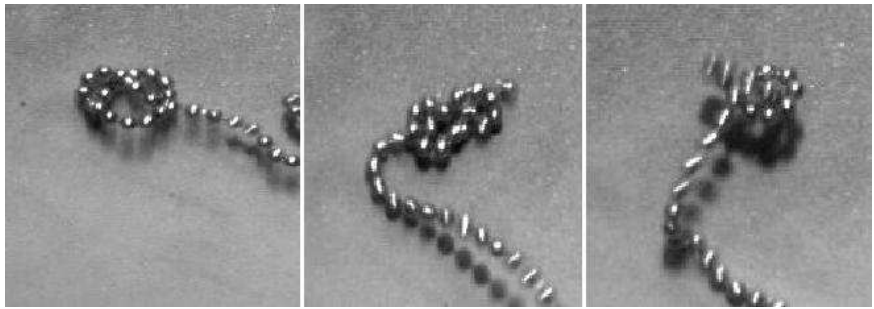}
\caption{Examples of knotting events, taking place at the 
end of the chain. 
\label{knot}}
\end{figure}
In Fig. \ref{total}, we plotted the cumulative distribution of
knotting times for the shorter chain, as obtained from the data 
of Fig. \ref{evolution}. The result is very well fitted by
an exponential distribution, using the mean knotting time $\tau_k$ 
from Table \ref{t1}. This is the distribution we will assume below
for both knotting and unknotting events, although the unknotting
distribution is more complicated \cite{BDVE01}. 

As a final result, we report the mean radius of gyration $<R_g>$ obtained 
by taking the spatial average of all beads of a chain, which was
done by image analysis using MATLAB. We then performed a temporal 
average, which we did separately for the knotted and the 
unknotted chain. As expected, $<R_g>$ is slightly larger for the 
unknotted state, but temporal fluctuations are prohibitively large
to make this a useful indicator of knottedness, as we had hoped
originally. The radius of gyration of course increases with chain
length, but at a slower rate than the linear extension of the
chain. 

\section{Theory and Discussion} 
Armed with the above observations, we can attempt a simple theory
for the probability of knots. We adopt a two-state description, in
which the chain is either in an unknotted or in a knotted state, the
probability of the latter being $P$. This can easily be generalized 
to allow for an arbitrary number of knots. Events are completely uncorrelated
in time, so the distribution of (say) unknotting times is assumed 
exponential, while reality is more complicated \cite{BDVE01}. Now
if the average unknotting time is $\tau$ as before, and the knotting 
time $\tau_k$, we obtain the following rate equation for 
$P$:
\begin{equation}
\dot{P} = (1-P)/\tau_k + P/\tau. 
\label{rate}
\end{equation}
The first term is the rate of knotting events, which only take
place if there is no knot. Conversely, the second term describes
the rate of unknotting. Equation (\ref{rate}) is solved very simply
with initial condition $P(0)=0$, giving
\begin{equation}
P(t) = \frac{1}{1+\tau_k/\tau}\left[1-\exp(-(\tau^{-1}+\tau_k^{-1})t)\right].
\label{p}
\end{equation}
for the probability after a time $t$. Now all that is needed are the
values of $\tau$ and $\tau_k$ as function of chain length.

For $\tau$ we take (\ref{BN}) as found in \cite{BDVE01}, which according
to Table \ref{t1} is consistent with our data. For the knotting time we take
\begin{equation}
\tau_k=\left\{\begin{array}{lcl}
            \infty  &\quad  & N<=N_{min}\\
            \tau_k^{(sat)} & \quad & N>N_{min}\\
             \end{array}\right. .
\label{unknot}
\end{equation}
As discussed above, this is based on the idea that knots are formed 
at the end of the chain. Once a length $N_{min}$, which gives sufficient 
freedom for knots to form, is exceeded, the rest of the chain no longer 
matters. By the same argument we also believe that the mean configuration
of the chain, as measured qualitatively by the radius of gyration,
is not of fundamental importance to calculate knotting probabilities. 
The constant $\tau_k^{(sat)}$ has to be determined empirically. 

For long chain lengths, $\tau$ becomes
large and (\ref{p}) reduces to $P=1-\exp(-t/\tau_k^{(sat)})$, 
independent of chain 
length, as expected. From the asymptotic value of $P=0.26$, taken from
Fig. \ref{comparison}, we deduce $\tau_k=100s$. For simplicity, we also 
assume that $N_{min}$ is essentially the same quantity as $N_0$ as identified
by \cite{BDVE01}, the length for which a knot falls out immediately. 
The result, (\ref{p}), is plotted as the solid line in Fig. \ref{comparison},
using (\ref{BN}),(\ref{unknot}). The agreement is quite good, considering there
is only one adjustable parameter. 

Ideally, even this adjustment could have been avoided, as $\tau_k$ could be 
taken directly from Table \ref{t1}. However, this gives about twice 
the value of $\tau_k$, which would lead to a significant disagreement in 
the asymptote of $P$. One simplifying assumptions of our model
was to disregard the {\it distribution} of knotting times. However
we suspect that the main reason for this disagreement lies in the difficulty
of preparing the chain in an unbiased fashion. At the beginning the ends of
the chain are excited to a higher degree, leading to a significant increase 
of the knotting probability. 

There is a significant body of work that remains to be done. 
Firstly, one would like to check our theory in greater detail,
by taking long time traces for a greater variety of chain lengths,
and measuring knotting probabilities for periods other than 30 s. 
Secondly, one would like to obtain a better description of the 
dynamics of the chain, and how it leads to knotting events. In 
particular, what sets the time scale for $\tau$ and $\tau_k$ ?
This could be addressed in part by changing the driving characteristics
of the plate, as well as the amount of confinement of the chain.
Preliminary experiments, performed in a container with side walls,
have in fact shown that boundary effects play a significant role
for the knotting probability. Finally, we have not considered the
probability for {\it different types} of knots, which are observed 
if the knotting time increases. 

\acknowledgments
This paper is the result of a Master's project at the University 
of Bristol physics department; an earlier version of the project
was carried out by C.J. Alvin and A. Barr. 
We are very grateful to the staff of BLADE at the University of 
Bristol Engineering department for hosting this experiment.
Clive Rendall and Tony Griffith in particular were instrumental in
setting up this experiment and gave their constant support. 
Nick Jones injected very useful ideas into the first phase of this project. 
One of us (JE) is grateful to Daniel Bonn and Jacques Meunier for their
warm welcome at the Laboratoire de Physique Statistique of the
ENS, Paris, where this paper was written.

\end{document}